\documentclass[twocolumn,twoside,slac_two]{revtex4}
\usepackage{graphicx}
\usepackage{fancyhdr}
\newcommand{\lsi}{LS~I~+61$^{\circ}$303}
\newcommand{\ls}{LS\,5039}

\begin{document}

\title{Results from the binaries \lsi\ \& \ls\ after 2.5 years of Fermi monitoring}


\author{D. Hadasch}
\affiliation{Institut de Ci\`encies de l'Espai (IEEC-CSIC),
              Campus UAB,  Torre C5, 2a planta,
              08193 Barcelona, Spain}
\author{on behalf of the \textit{Fermi}-LAT collaboration}

\begin{abstract}

The Fermi Large Area Telescope (LAT) has made the first definitive GeV detections
of the binaries \lsi\ and \ls\ in the first year after its launch
in August 2008. These detections were unambiguous because, apart from a reduced positional uncertainty,
the $\gamma$-ray emission in each case was orbitally modulated with the corresponding orbital period.
The LAT results posed new questions about the nature of these objects,
after the unexpected observation of an exponential cutoff in the GeV $\gamma$-ray spectra of both
LS I +61$^\circ$303 and LS\,5039, at least along part of their orbital motion.
We present here the analysis of new data from the LAT, comprising 2.5 years of observations
through which LS I +61$^\circ$303 continues to provide some surprises. We find
a sudden increase in flux in March 2009 and a steady decrease in the flux fraction
modulation.
The LAT now detects emission up to
30\,GeV, where prior datasets led to upper limits only.
At the same time, contemporaneous TeV observations either no longer detected the source, or found it -at least in some orbits-
close to periastron, far from the usual phases in which the source usually appeared at TeV energies.
The on-source exposure
of LS\,5039 has also drastically increased along the last years,
and whilst our analysis shows no new
behavior in comparison with our earlier report, the higher statistics of the current dataset allows for a deeper investigation of its
orbital and spectral evolution.
\end{abstract}

\maketitle

\thispagestyle{fancy}


\section{Introduction}	

To date there are only five X-ray binaries that have been detected at high (HE;
0.1-100\, GeV) or very high-energies (VHE; $>$100 GeV): \lsi\ \cite{albert2006,acciari2008,abdo2009a}; \ls\ \cite{aharonian2005b,abdo2009b}, PSR
B1259-63 \cite{aharonian2005a}, Cyg X-3 \cite{abdo2009}, and Cyg X-1 \cite{albert2007,sabatini2010}. Of these sources, only \lsi\
and \ls\ exhibit a non-transient behavior at high-energies and consequently, have been detected as persistent, albeit variable,
sources of $\gamma$-ray emission. They also share the property of being (together with PSR B1259-63, see \cite{abdo2010})
sources detected at both GeV and TeV energies. The other systems mentioned have been unambiguously detected only in one band,
either at GeV or at TeV.

The early LAT reports of GeV emission from \ls\ and \lsi\ were
based upon 6--9 months of survey observations \cite{abdo2009a,abdo2009b}. Both sources were detected
at high significance and were unambiguously identified with the binaries by their flux modulation at the corresponding orbital periods,
26.4960 for \lsi\ \cite{gregory2002} and 3.90603 days for \ls\ \cite{casares2005}. The modulation
patterns were consistent with expectations from inverse Compton scattering models, and
were anti-correlated in phase with pre-existing TeV measurements (e.g., \cite{albert2009,aharonian2006}).
The anti-correlation of GeV--TeV fluxes is in fact a generic feature embedded
in inverse Compton models describing the TeV fluxes, where the GeV emission is
enhanced (reduced) when the highly relativistic electrons seen by the observer
encounter the seed photons head-on (rear-on); (e.g., \cite{boettcher2005,bednarek2007}).
Fermi measurements provided a generic confirmation for
these inverse Compton models.

Both sources presented exponential cutoffs in their high-energy spectra, at least along part of the orbit. To be precise, an exponential cutoff was a better fit
to the SED --as compared with a pure power-law-- in phases surrounding the
inferior conjunction (INFC) of \ls, and when taking into account the average
spectrum of \lsi\ along its whole orbit. Statistical limitations with the used
amount of data prevented the determination or the ruling out of an exponential
cutoff in any orbital cut of \lsi\ or in the superior conjunction (SUPC) of
\ls. The spectral energy distributions with the exponential cutoffs that were
reported  were reminiscent of the many pulsars the LAT has discovered since
launch \cite{abdo2010b}, although this was far from a proof of an \ls\ or
\lsi\ pulsar nature. No pulsations has been found at GeV energies.

In this work we present the results of the analysis of 2.5 years of LAT survey observations of both LS I +61$^\circ$303 and
LS\, 5039. We investigate any long-term flux variations of the sources and explore too the possible spectral variability for both systems. 
We expect that our analysis will constrain future theoretical studies on these interesting objects.

\section[data]{Observations and data reduction}\label{data}

The Fermi Gamma-ray Space Telescope, launched on 2008 June 11, carries onboard
the Large Area Telescope (LAT).
The LAT  is an electron-positron pair production
telescope, featuring solid state silicon trackers and cesium iodide
calorimeters, sensitive to photons from $\sim$20\,MeV to $>$300\,GeV \cite{atwood2009}.
It has a large field of view with $\sim$2.4 sr (at 1\,GeV) and an effective area of $\sim$8000\,cm$^2$ for $>$1\,GeV.

The Fermi survey mode operations began on 2008 August 4. The full dataset used for this analysis spans 2008 August 4, through 2010 December 4.
The data were reduced and analyzed using the {Fermi Science Tools v9r20 package}.\footnote[1]{See the Fermi Space Science Centre (FSSC) website for details of the Science Tools:
http://fermi.gsfc.nasa.gov/ssc/data/analysis/}
The standard
onboard filtering, event reconstruction, and classification were applied to the data \cite{atwood2009}. The high-quality ``diffuse" event class was used. Throughout the
analysis the ``Pass 6 v3 Diffuse" instrument response functions (IRFs) were applied.
%Time periods when the target source was observed at a zenith angle greater than 105$^\circ$ and for observatory
%rocking angles of greater than 43$^\circ$ were excluded to avoid contamination from Earth albedo
%photons. 
Where required in the analysis, models for the Galactic diffuse emission (\textit{gll\_iem\_v02.fit})
and isotropic backgrounds (\textit{isotropic\_iem\_v02.txt}) were used.\footnote[2]{Descriptions of the models are available from the FSSC: http://fermi.gsfc.nasa.gov/ssc}

The binned maximum-likelihood method of {\tt gtlike}, which is part of the
ScienceTools, was applied to determine the intensities and spectral parameters
presented in this paper.
We used all photons with energy $>$100\,MeV in a circular region of interest (ROI) of $10^{\circ}$ radius centered at the position of \lsi\ and \ls, respectively.
For source modeling, the 1FGL catalog \cite{abdo2010a}, derived from 11 months of survey data, was taken.
%The energy spectra of point sources included in the catalog within our ROI are modeled by simple power-laws
%with the exception of known $\gamma$-ray pulsars, which were
%modeled by power-laws with exponential cutoffs.
%The spectral parameters were fixed to the catalog values except for the sources within 3 degrees of the candidate location. For these latter sources, the flux normalization was left free. Obviously, all spectral parameters of the two subject binaries were left free for the fit too.
%The folded lightcurves were derived by performing {\tt gtlike} fits for each phase bin.

\section{LS\,5039 Results}\label{LS5039}

LS~5039 is located in a complicated region toward the inner Galaxy with high Galactic diffuse emission and many gamma-ray sources.
In an earlier publication about this source \cite{abdo2009b} we derived that a power law plus an exponential cutoff describes best the data.
The photon index was $\Gamma = 1.9 \pm 0.1_{\mathrm{stat}} \pm 0.3_{\mathrm{syst}}$; the 100\,--300\,GeV flux was $(4.9 \pm 0.5_{\mathrm{stat}} \pm 1.8_{\mathrm{syst}}) \times 10^{-7} \mathrm{ph\, cm^{-2}\, s^{-1}}$
and the cutoff energy was found to be $2.1 \pm 0.3_{\mathrm{stat}} \pm 1.1_{\mathrm{syst}}\, \mathrm{GeV}$.

Firstly, we analyzed the orbitally-averaged data of LS~5039. As spectral models for LS~5039, we used a power law
as well as a power law with an exponential cutoff, and compared likelihood obtained to test the significance of
a spectral cutoff.
The likelihood ratio between the power-law and cutoff power-law cases clearly indicates that the power-law assumption is rejected.

\begin{figure*}[ht]
    \begin{center}
    \includegraphics*[width=0.49\textwidth,angle=0,clip]{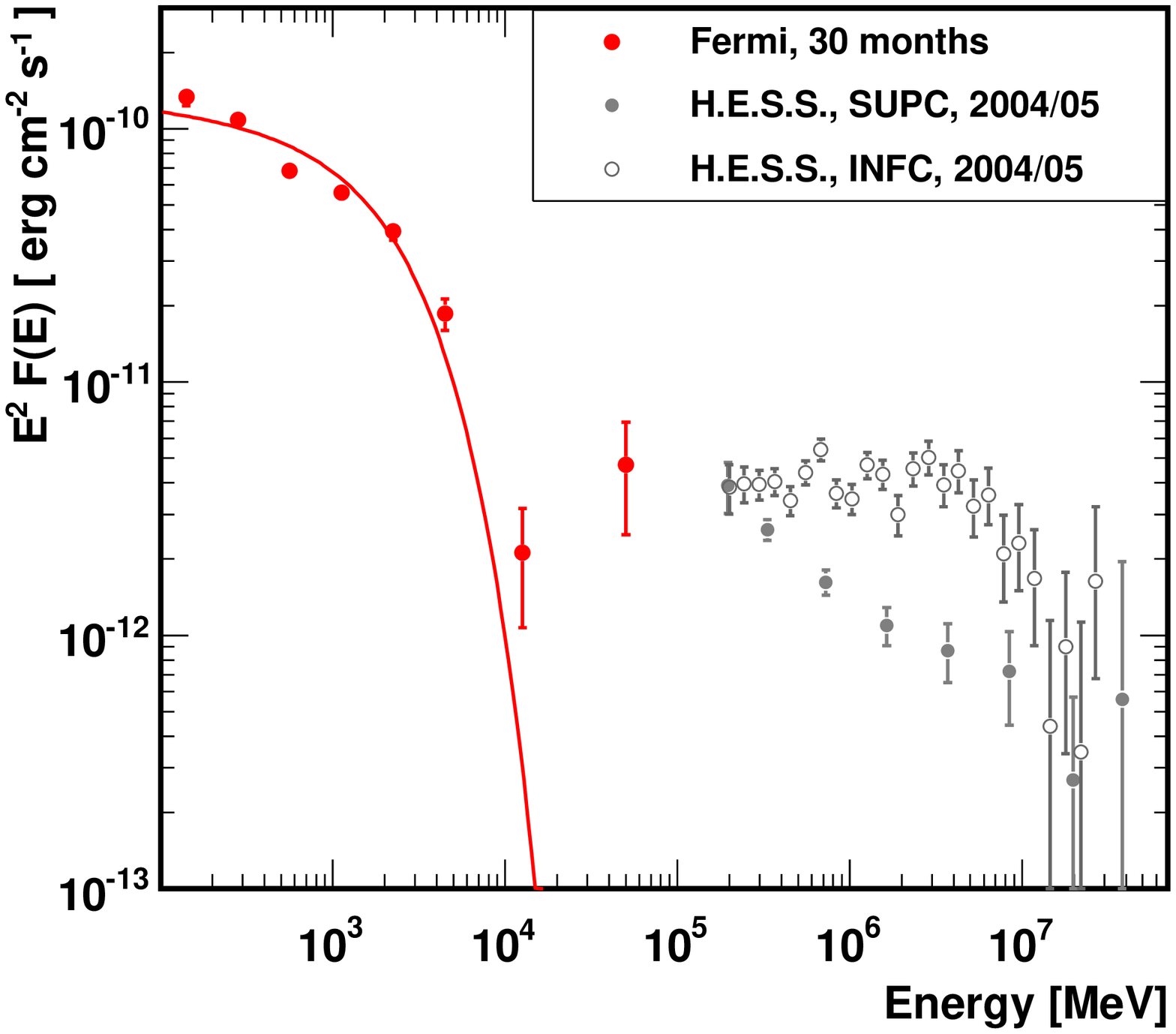}
    \includegraphics*[width=0.49\textwidth,angle=0,clip]{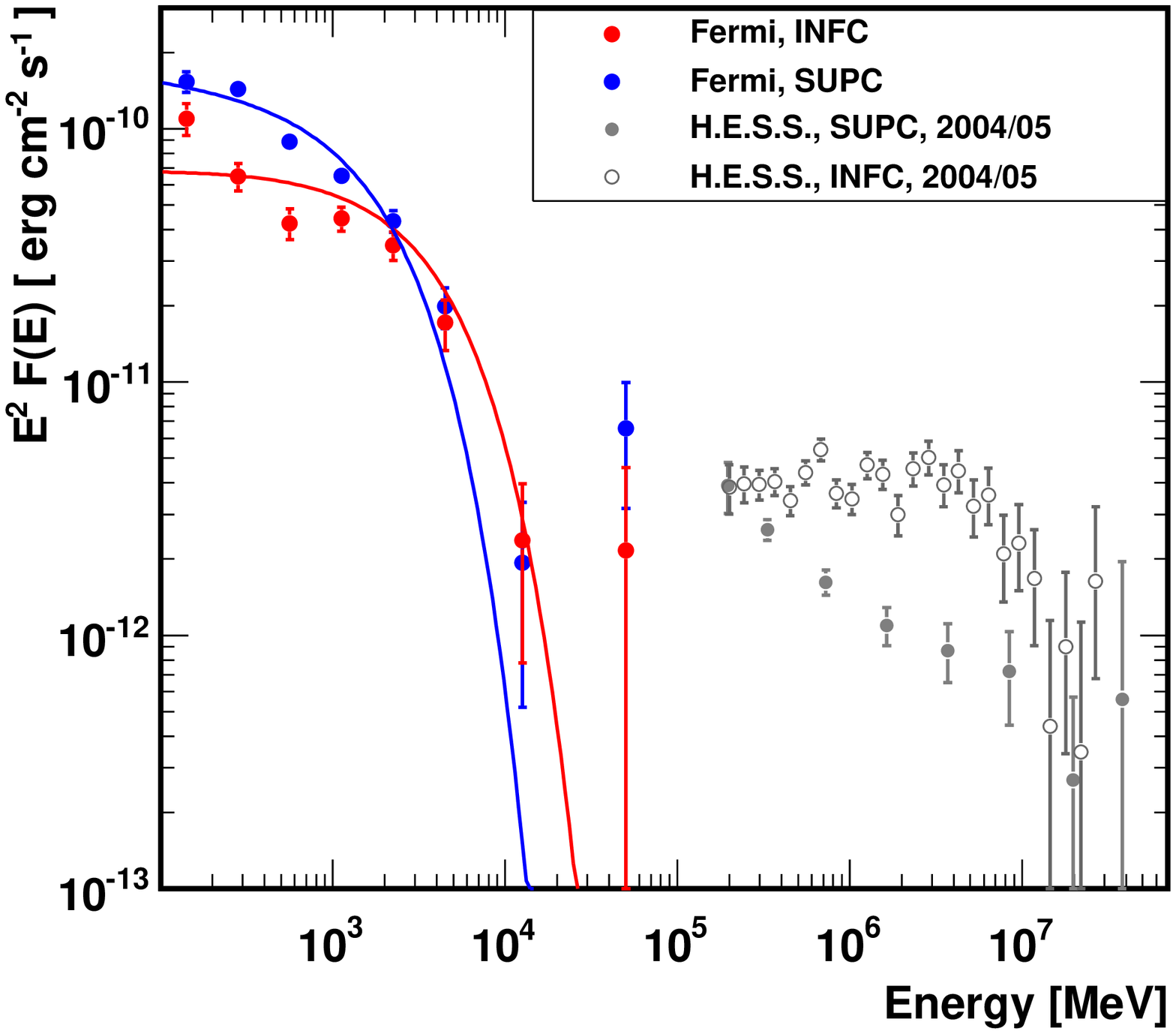}
\caption{\label{fig:LS5039_spec}\textit{Left: }The overall spectrum of \ls\ over 2.5 years of data are shown in red.
Also, very high-energy data points from H.E.S.S. are plotted, but one has to keep in mind that they are not simultaneous. The gray filled points correspond to inferior conjunction of the system and the open ones to superior conjunction.
\textit{Right: }Spectra for inferior (red) and superior (blue) conjunction of the system. Note that all plots are preliminary.}
    \end{center}
\end{figure*}

Spectral points at each energy band were obtained by dividing the data set into each energy bin and
performing maximum likelihood fits for each of them. Resulting spectral energy distribution (SED) is plotted
in Figure~\ref{fig:LS5039_spec}. 

Following the H.E.S.S. analysis by  \cite{aharonian2006} and our previous analysis,
the whole data set was divided into two orbital intervals:
superior conjunction (SUPC; $\phi < 0.45$ and $0.9 < \phi$) and inferior conjunction (INFC; $0.45 < \phi < 0.9$).
The SUPC and INFC data were analyzed in the same way as the orbitally averaged data.
Being consistent with our previous paper, the power-law assumption for the SUPC spectrum is
rejected.
Although a single power law was not rejected for INFC in our previous analysis using 10-month data,
a cutoff power law is preferred also for INFC with the present data set.
The corresponding SED is shown in Figure~\ref{fig:LS5039_spec}.

\section{LS I +61$^\circ$303 Results}\label{LSI}

%In our earlier publication \cite{abdo2009a}, we have reported that the orbitally averaged (i.e., without any orbital cuts)
%LAT data of LS I +61$^\circ$303 from August 2008 through February 2009 were well fitted by a power-law
%plus an exponential cutoff.
%
%The photon index was found to be $\Gamma = 2.21 \pm 0.04_{\mathrm{stat}} \pm 0.06_{\mathrm{syst}}$; the flux above
%100 MeV was $(0.82 \pm 0.03_{\mathrm{stat}} \pm 0.07_{\mathrm{syst}}) \times 10^{-6} \mathrm{ph\, cm^{-2}\, s^{-1}}$ and the cutoff energy was
%$6.3 \pm 1.1_{\mathrm{stat}} \pm 0.4_{\mathrm{syst}}\, \mathrm{GeV}$.
%
%The currently obtained spectral points and the best-fit averaged over all the orbital phases of the \lsi\
%system are shown in the left panel of Figure \ref{fig:whole_data}, together with the previously derived results
%from the LAT and the higher-energy, non-simultaneous data points obtained by the Cherenkov telescope experiments.
%As one can see there, the LAT data along the whole orbit are still best described by a power-law with an exponential
%cutoff. 
%The relative TS value \cite{mattox} comparing a fit with a power-law and a fit with a power-law with an exponential cutoff clearly favors the latter.
%We have also tested a broken power-law shape as an a priori assumption to fit the data, but a power-law with exponential cutoff is still prefered.

%\subsection{Average flux change around March 2009}\label{change}

\begin{figure*}[ht]
    \begin{center}
   \includegraphics*[width=0.32\textwidth,angle=0,clip]{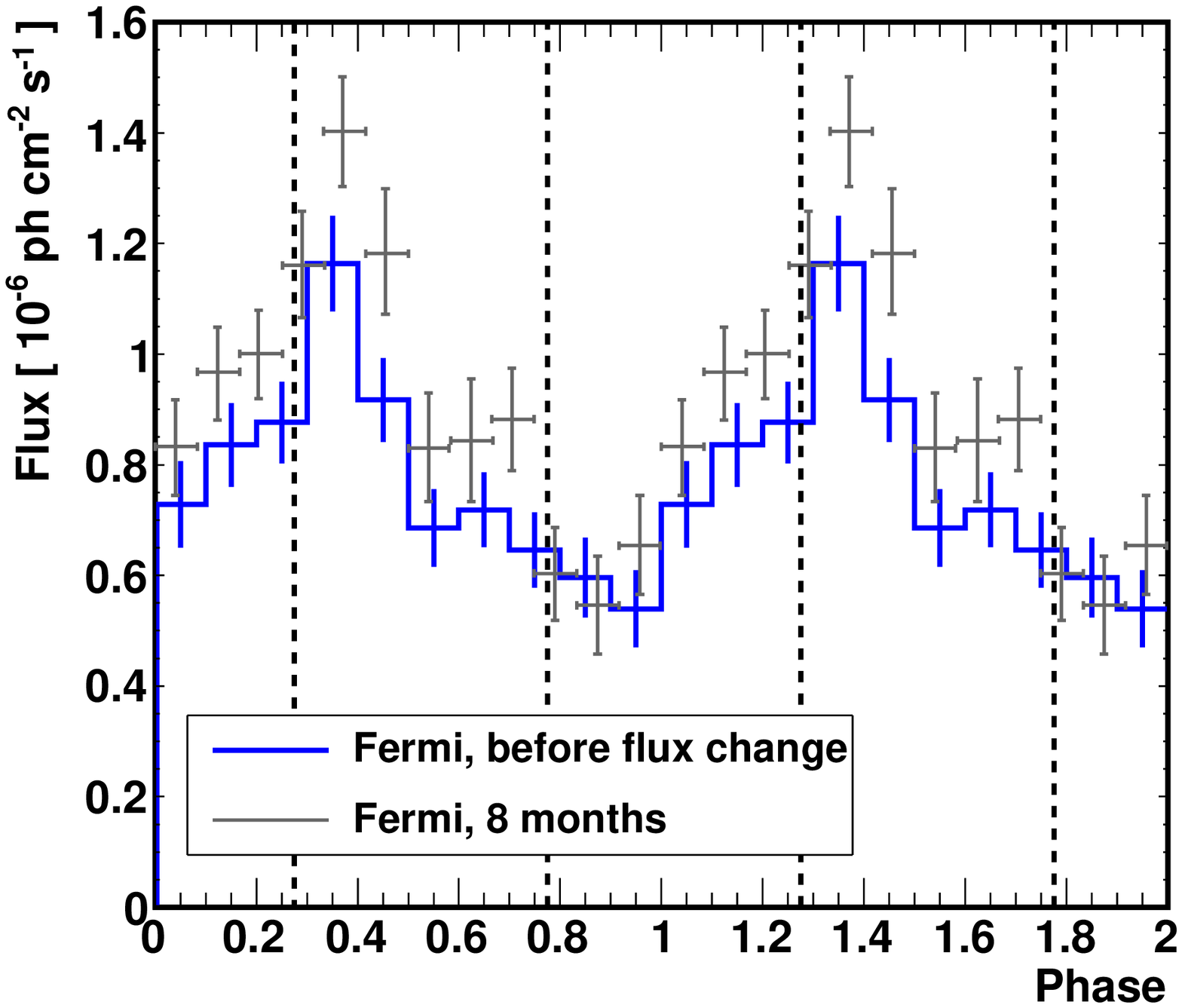}
    \includegraphics*[width=0.32\textwidth,angle=0,clip]{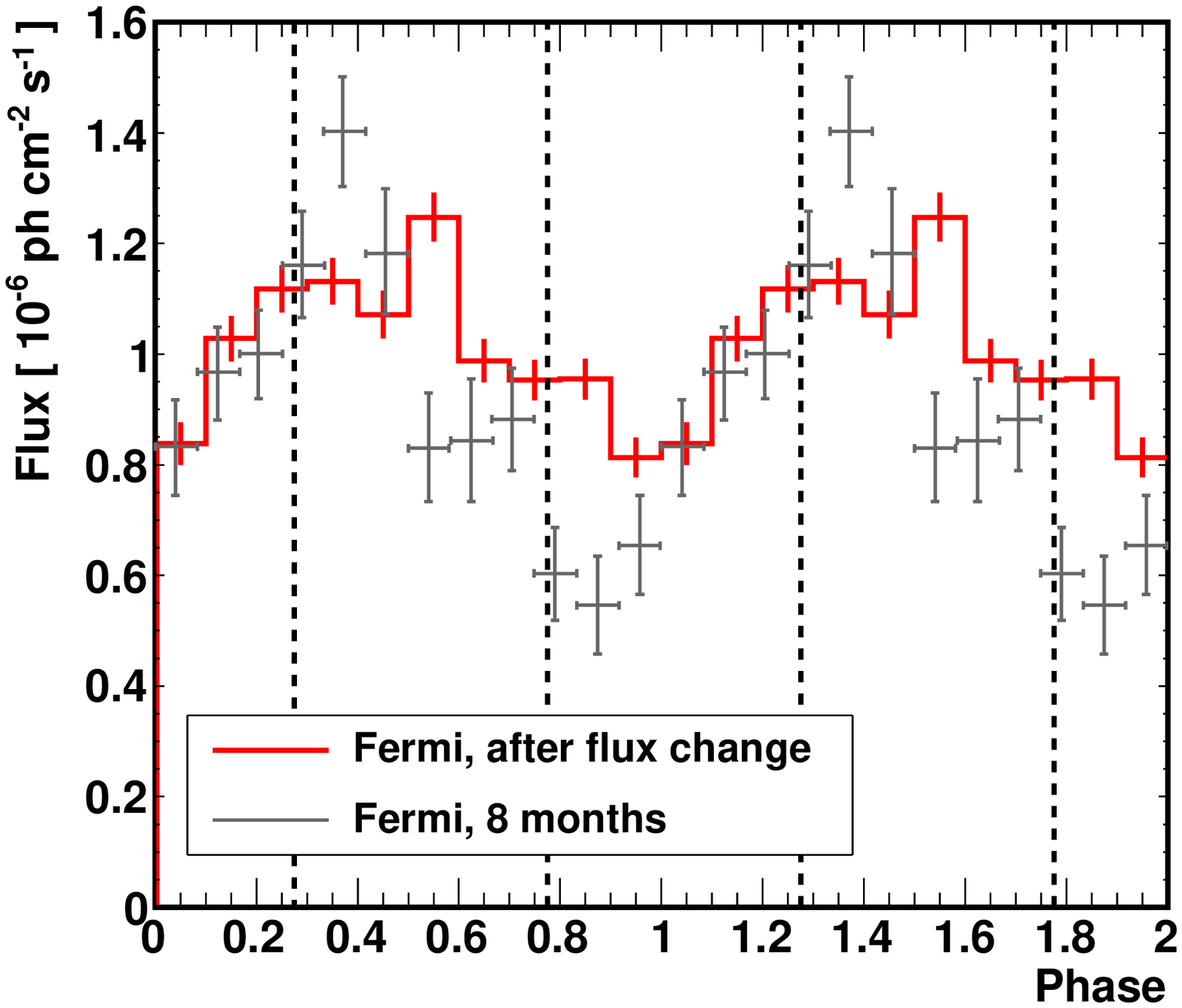}
\includegraphics*[width=0.32\textwidth,angle=0,clip]{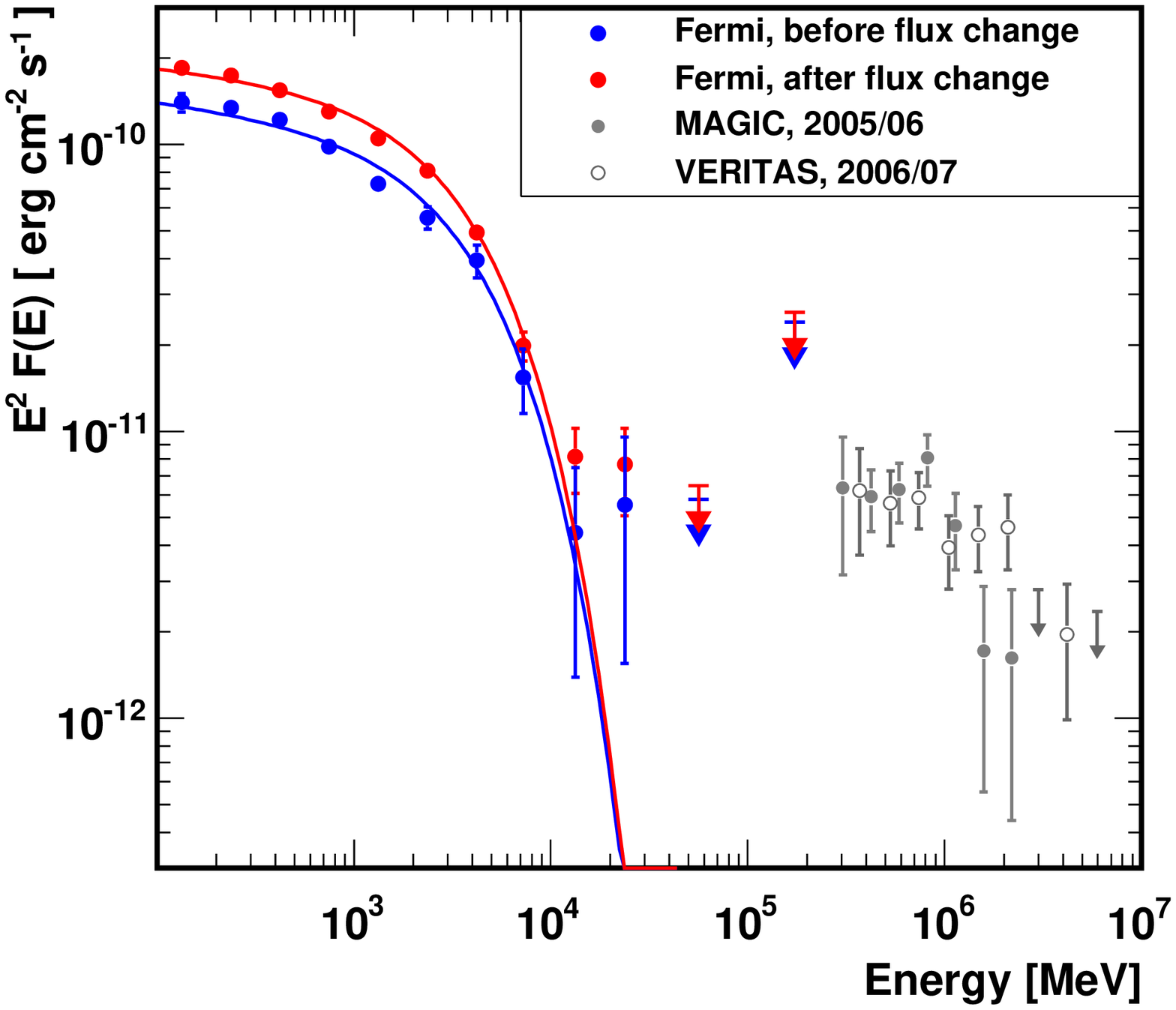}
\caption{\label{fig:before_after} \textit{Left: }Folded lightcurve of \lsi\ before the flux change (blue), when the modulation is still clearly visible. \textit{Middle: }Folded lightcurve after the flux change in March 2009 (red) when the modulation becomes less pronounced. The modulation gets fainter.  \textit{Right: } Comparison of the spectra derived before (blue) and after (red) the flux change in March 2009. Note that all plots are preliminary.}
    \end{center}
\end{figure*}

LS I +61$^\circ$303 is one of the brightest sources in the $\gamma$-ray sky and towers above  other
emitters in its neighborhood. This has allowed us to clearly detect, in March 2009, a $\sim$30\% increase on the
average flux.
We graphically show the flux  change in Figure~\ref{fig:before_after}, by plotting the folded lightcurves before and after the found transition
in March 2009.
Before the transition, the modulation was clearly seen and is compatible with the already published phasogram, whereas afterwards, the modulation gets fainter. Note that
the datasets corresponding to the reported results \cite{abdo2009a} and what we here referred to as {\it before the flux change}
span almost exactly the same time range, with the consequence of our current analysis essentially reproducing the one previously published\cite{abdo2009a}. 
The spectra derived before and after this flux change are also shown in Figure~\ref{fig:before_after}, where the increase in flux is obvious.

\begin{figure*}[ht]
    \begin{center}
\begin{minipage}{0.49\textwidth}
    \includegraphics*[width=0.99\textwidth,angle=0,clip]{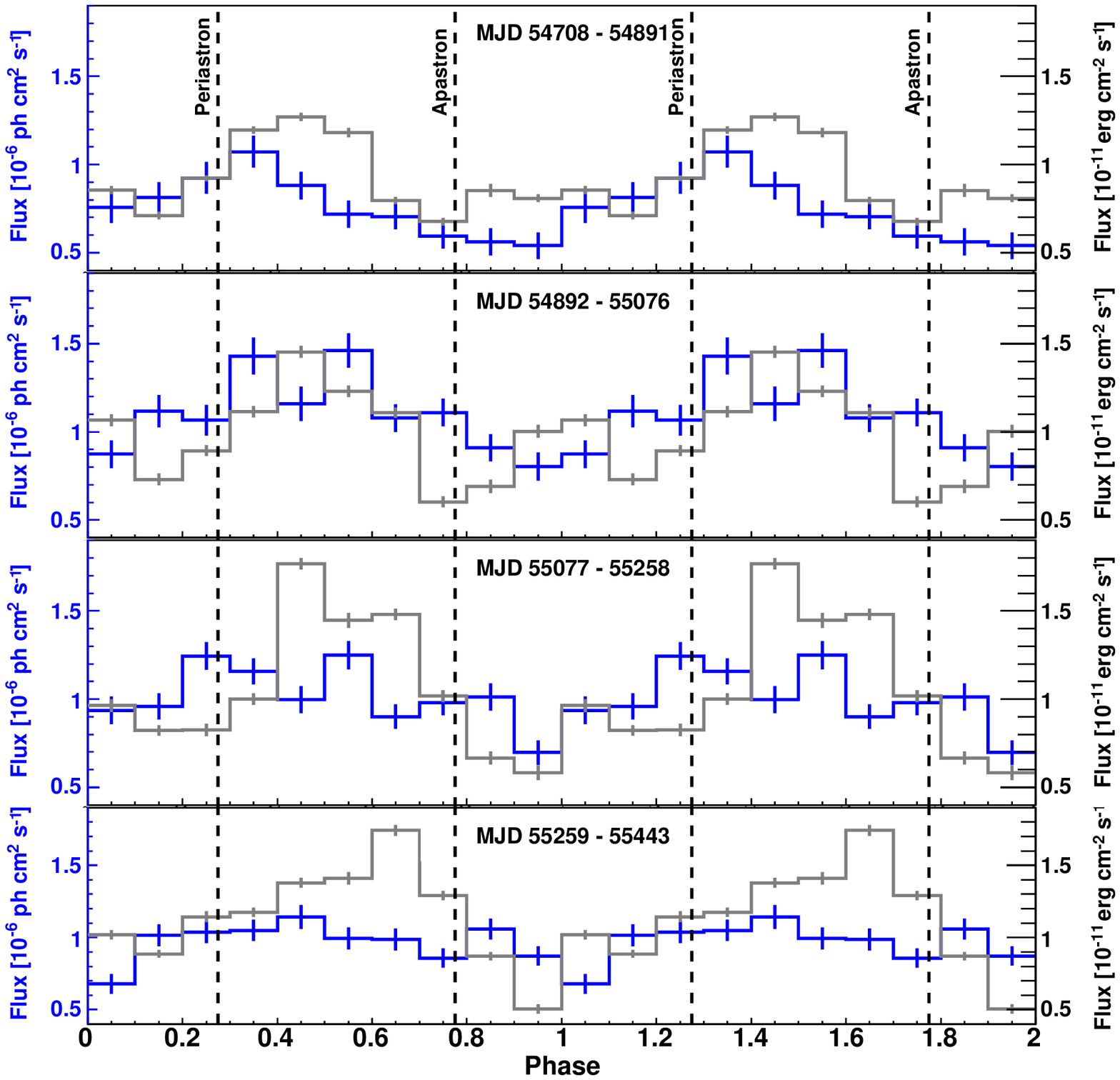}
\caption{\label{fig:multi}Comparison between the $\gamma$-ray (blue) and the X-ray (gray) data of \lsi\ . For each of the 4 separate 6-months periods, the 3-10\,keV (gray) and the 100\,MeV-300\,GeV (blue) folded lightcurves are shown.}
\end{minipage}
\begin{minipage}{0.49\textwidth}
    \includegraphics*[width=0.99\textwidth,angle=0,clip]{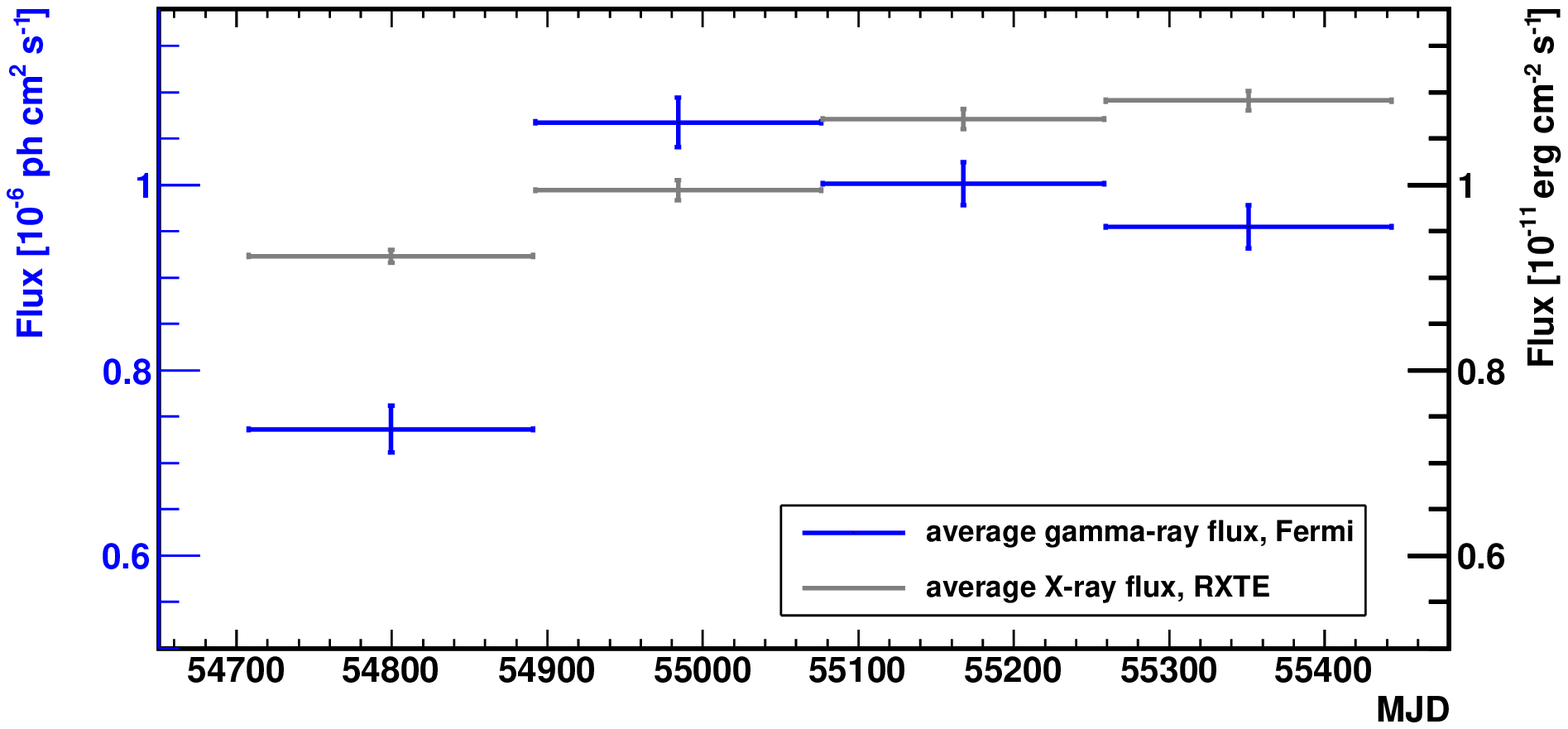}
\caption{\label{fig:average_flux}The average flux in X- (gray) and $\gamma$-rays (blue) are shown. Note that all plots are preliminary.}
\end{minipage}
    \end{center}
\end{figure*}

\subsection{The multi-wavelength context}

\lsi\ has also been monitored with the {\it RXTE}--Proportional Counter Array (PCA) and folded lightcurves
were produced. In the left panel of Figure
\ref{fig:multi}, we show a direct comparison between the phasograms in X-ray
and in $\gamma$-rays, with simultaneously taken data. We divide the whole LAT
dataset into 4 periods of 6 months each and compare them with correspondingly
obtained PCA X-ray data. The division in periods of 6-months is justified in
order to have enough statistics in $\gamma$-rays for each individual time bin,
and such that orbit-to-orbit X-ray variability does not dominate the flux
fraction changes.
It is apparent that the X-ray
modulation is always visible in each of these five panels, albeit with variable amplitude of flux modulation. Instead, at GeV energies,
the LAT data indicates that the modulation fades away until the variability along the orbit is barely visible in the last 6 months of our data.
In Figure \ref{fig:average_flux} we show the average $\gamma$- and X-ray fluxes fitted in each of the periods considered. The X-ray flux is increasing, whereas the $\gamma$-ray flux seems to decrease again after the flux change in March 2009. 

\section{Summary}

After two years of continuous data taking of the two binary systems \ls\ and
\lsi\ we confirm that their energy spectra are best described by a power law function 
with an exponential cutoff.
For \ls\,, in comparison with earlier publications of this source, with larger data set at hand 
we can extend the observed spectrum to higher energies.  
No significant emission of \lsi\ above 30\,GeV has been detected with the $Fermi$-LAT yet.
Furthermore, in the case of \lsi\, variable modulation in the folded light curves was observed.
Before a flux change in March 2009, the modulation was clearly visible, whereas afterward it faints away.
This behavior was not predicted by any theoretical model so far and has to be investigated further.

\bigskip % extra skip inserted
\begin{acknowledgments}
The Fermi LAT Collaboration acknowledges generous ongoing support from a number of
agencies and institutes that have supported both the development and the operation of the LAT as
well as scientific data analysis. These include the National Aeronautics and Space Administration
and the Department of Energy in the United States, the Commissariat \`a l'Energie Atomique and
the Centre National de la Recherche Scientifique / Institut National de Physique Nucl\'eaire et de
Physique des Particules in France, the Agenzia Spaziale Italiana and the Istituto Nazionale di Fisica
Nucleare in Italy, the Ministry of Education, Culture, Sports, Science and Technology (MEXT),
High Energy Accelerator Research Organization (KEK) and Japan Aerospace Exploration Agency
(JAXA) in Japan, and the K. A. Wallenberg Foundation, the Swedish Research Council and the
Swedish National Space Board in Sweden.
Additional support for science analysis during the operations phase is gratefully acknowledged
from the Istituto Nazionale di Astrofisica in Italy and the
Centre National d'\'Etudes Spatiales in France.
This work has been additionally supported by the Spanish CSIC and MICINN and the Generalitat de Catalunya, through grants AYA2009-07391 and SGR2009-811, as
well as the Formosa Program TW2010005. SZ acknowledges supports from  National Natural Science Foundation of
China (via NSFC-10325313, 10521001, 10733010, 10821061 and 11073021),
and 973 program 2009CB824800.
\end{acknowledgments}

\bigskip % extra skip inserted
%\begin{thebibliography}{9}   % Use for  1-9  references
%\begin{thebibliography}{99} % Use for 10-99 references

\end{document}